\documentclass[conference]{IEEEtran}
\IEEEoverridecommandlockouts

\usepackage{cite}
\usepackage{amsmath,amssymb,amsfonts}
\usepackage{algorithmic}
\usepackage{graphicx}
\usepackage{textcomp}
\usepackage{xcolor}
\def\BibTeX{{\rm B\kern-.05em{\sc i\kern-.025em b}\kern-.08em
    T\kern-.1667em\lower.7ex\hbox{E}\kern-.125emX}}

\newcommand{\size}[2]{{\fontsize{#1}{0}\selectfont#2}}

\usepackage{booktabs}
\usepackage{outlines}
\usepackage{epsfig}

\PassOptionsToPackage{hyphens}{url}
\usepackage[hyphens]{url}
\usepackage[bottom,para]{footmisc}
\usepackage{multirow}
\usepackage{caption}
\usepackage{xspace} %
\usepackage{lipsum}
\usepackage{subcaption}
\usepackage{amsfonts} %
\DeclareGraphicsExtensions{.png, .pdf, .jpg, .bmp}

\usepackage{filecontents}
\usepackage{tikz}

\usepackage{tabularx}
\usepackage{multirow}
\usepackage{balance}
\usepackage{makecell}
\usepackage{adjustbox}

\usepackage{tcolorbox}

\usepackage{enumitem}
\setlist{topsep=1pt,itemsep=1pt,parsep=1pt,itemindent=0pt,leftmargin=0.13in}

\usepackage{soul}
\soulregister\cite7
\soulregister\ref7

\usepackage{fontawesome} %

\newcommand{\Tm}{\faTimes}
\newcommand{\Ck}{\faCheck}

\newcommand{\ie}{{\em i.e., \/}}
\newcommand{\eg}{{\em e.g., \/}}
\newcommand{\etc}{{\em etc. \/}}

\newcommand{\hide}[1] {}

\usepackage{booktabs} %
\usepackage{listings}
\usepackage{color}

\usepackage{url}

\definecolor{dkgreen}{rgb}{0,0.6,0}
\definecolor{gray}{rgb}{0.5,0.5,0.5}
\definecolor{mauve}{rgb}{0.58,0,0.82}

\newcommand\blfootnote[1]{%
  \begingroup
  \renewcommand\thefootnote{}\footnote{#1}%
  \addtocounter{footnote}{-1}%
  \endgroup
}

\newcommand{\Scut}[1] {}

\begin{document}

\title{Analyzing Open-Source Serverless Platforms:\\ Characteristics and Performance}

\author{\IEEEauthorblockN{
Junfeng Li\textsuperscript{1},
Sameer G. Kulkarni\textsuperscript{2},
K. K. Ramakrishnan\textsuperscript{3},
Dan Li\textsuperscript{1}
}
\IEEEauthorblockA{
\textsuperscript{1}Tsinghua University,
\textsuperscript{2}IIT, Gandhinagar,
\textsuperscript{3}University of California, Riverside
\\
Email: hotjunfeng@163.com, 
sameer.sameergk@gmail.com,
kk@cs.ucr.edu,
tolidan@tsinghua.edu.cn
\vspace{-8mm}
}

\thanks{
	In this paper, we significantly add the following work based on a previous workshop version~\cite{wosc} published at WoSC'19: 
	(1) We describe the salient characteristics of four serverless platforms in more detail in \S\ref{sec:platform-characteristics}, including new figure illustration;
	(2) We elaborate resource/workload-based auto-scaling frameworks in \S\ref{sec:platform-characteristics};
	(3) We evaluate four serverless frameworks with more experiments in \S\ref{sec:evaluation};
	we also add detailed analysis about the root cause of performance differences among these serverless frameworks in \S\ref{sec:evaluation};
	(4) We provide insights and future work for serverless platforms in \S\ref{sec:insights}.
}

}

\maketitle

\begin{abstract}
Serverless computing is increasingly popular because of its lower cost and easier deployment.
Several cloud service providers (CSPs) offer serverless computing on their public clouds, but it may bring the vendor lock-in risk. To avoid this limitation, many open-source serverless platforms come out to allow developers to freely deploy and manage functions on self-hosted clouds. 
However, building effective functions requires much expertise and thorough comprehension of platform frameworks and features that affect performance. It is a challenge for a service developer to differentiate and select the appropriate serverless platform for different demands and scenarios.
Thus, we elaborate the frameworks and event processing models of four popular open-source serverless platforms and identify their salient idiosyncrasies.
We analyze the root causes of performance differences between different service exporting and auto-scaling modes on those platforms.
Further, we provide several insights for future work, such as auto-scaling and metric collection.

\blfootnote{\newline This is the author's preprint of the work (updated on May 28th, 2021). It is posted here only for your personal use. Not for redistribution. The official version is: \url{https://ksiresearch.org/seke/seke21paper/paper129.pdf}}
\blfootnote{\newline DOI reference number: 10.18293/SEKE2021-129}
\end{abstract}

\begin{IEEEkeywords}
cloud computing, serverless, function-as-a-service, characteristic, performance
\end{IEEEkeywords}

\size{9}{\textbf{\textit{Reference Format---}}}
\size{9}{Junfeng Li, Sameer G. Kulkarni, K. K. Ramakrishnan, Dan Li. Analyzing Open-Source Serverless Platforms: Characteristics and Performance. In \textit{The 33rd International Conference on Software Engineering and Knowledge Engineering (SEKE '21)}, 2021. \url{https://doi.org/10.18293/SEKE2021-129}}

\section{Introduction}
Serverless computing has ushered in a new era in cloud computing. 
Cloud computing 
seeks to provide computing and storage services at large scale and low cost to end-users through economies of scale and effective multiplexing~\cite{li2018towards}.    
Serverless computing puts multiplexing and scalability to the next level by allowing providers to commit just the required amount of resources to a particular application 
and utilize the resources for just the time needed to execute an invoked function. Resources are scaled dynamically to meet the demand of user requests.
Unlike traditional cloud deployment models%
, where a number of computing instances are deployed well in advance,
serverless computing achieves nearly zero resource cost when there is no demand, and scales to as many instances as needed to meet the traffic demand. Thus, serverless computing could be both scalable and cost effective. 

In addition to scalability and multiplexing, serverless computing allows developers to build, deploy and run the application on demand without focusing on server management, according to the Cloud Native Computing Foundation (CNCF)~\cite{yaroncncf}.  
When an event is triggered, a piece of infrastructure is allocated dynamically for function execution.
The underlying details of resource management, \ie resource allocation, data transmission and function execution, are decoupled from the user.
Many cloud service providers (CSPs) offer serverless computing platforms on their public clouds, such as Amazon Web Services (AWS) Lambda, which
is an event-driven serverless platform that enables to implement and deploy application in any supported languages %
and execute on-demand as docker containers. 
Since public serverless platforms may incur vendor lock-in risk, many open-source serverless platforms spring up and allow developers to freely deploy and manage functions on self-hosted clouds.
However, building effective functions requires much expertise and in-depth understanding of platform frameworks and characteristics that affect performance. It is a challenge for a service developer to differentiate and select the proper serverless platform in different scenarios.

To help developers choose suitable open-source platforms to deploy efficient services, we systematically identify and analyze the salient characteristics of several popular open-source serverless platforms (\ie Knative\footnote{\url{https://github.com/knative}}, Kubeless\footnote{\url{https://kubeless.io}}, Nuclio\footnote{\url{https://nuclio.io}} and OpenFaaS\footnote{\url{https://www.openfaas.com}}) and compare their performance.
Our key contributions include:
\begin{itemize}
    \item We provide an understanding of the platform frameworks and interaction between different components of four popular open-source serverless platforms. 
    \item We analyze the salient features of each platform, such as the built-in workload-based auto-scaling mechanism and the event processing model inside the function pod. 
    \item We evaluate the performance of different service exporting and auto-scaling modes, and analyze the root cause of performance gap among different serverless platforms. 
    \item We give several insights for future work, such as auto-scaling and metric collection.
\end{itemize}

\section{Background}
\begin{table*}[h]
    \centering
    \setlength{\extrarowheight}{0.12 cm}
    \begin{tabular}{|c||c|c|c|c|}
        \hline
        \textbf{Feature}                                                      & \textbf{Nuclio}          & \textbf{OpenFaaS}            & \textbf{Knative}             & \textbf{Kubeless}     \\ \hline \hline
        \textbf{Queue inside Function Pod}                                    & \Ck                      & \Ck                          & \Ck                          & \Tm                   \\ \hline
        \textbf{\begin{tabular}[c]{@{}c@{}}Support for Multiple Workers \\ 
        in Function Pod\end{tabular}}                                         & \Ck                      & \Tm                          & \Ck                          & \Tm                   \\ \hline
        \textbf{Function Startup Policy}                                      & Warm Start               & Cold/Warm Start              & Cold/Warm Start              & Cold Start            \\ \hline
        \textbf{Service Export Method}                                        & Ingress Gateway/NodePort & API Gateway/Ingress Gateway  & Ingress Gateway              & Ingress Gateway       \\ \hline
        \textbf{Runtime Metric Collection}                                    & Metric Server            & Metric Server/API Gateway    & Metric Server/Queue-proxy    & Metric Server         \\ \hline
        \textbf{Auto-scaling Mode}                                            & CPU/Memory               & CPU/Memory/RPS               & CPU/Memory/Concurrency/RPS   & CPU/Memory            \\ \hline
        \textbf{Scale-to-zero}                                                & \Tm                      & \Tm                          & \Ck                          & \Tm                   \\ \hline
    \end{tabular}
    \caption{Comparision of popular open-source serverless platforms}
	\label{table:platform-comparison}
    \vspace{-3mm}
\end{table*}

Many cloud service providers (CSPs) offer serverless computing platforms on their public clouds, such as Amazon Web Services (AWS) Lambda, Google Cloud Functions, Azure Functions and Alibaba Cloud Function Compute.
The developers are required to design and deploy their serverless functions based on the supporting services provided by CSPs, such as message queuing, storage and database. Thus it incurs the risk of vendor lock-in.
The deployed serverless functions rely heavily on specific CSPs, and it is difficult to migrate existing functions to either self-hosted clusters or other public clouds.

The open-source serverless platforms bring more flexibility and allow developers to freely deploy and manage functions on self-hosted clouds. 
However, there are still some challenges for open-source serverless platforms: 
(1) it requires a deep understanding of platform features to build effective functions;
(2) the developers should manage and maintain serverless platforms by themselves, which requires much expertise of platform frameworks and infrastructures;
(3) the performance of open-source serverless platforms may vary in different scenarios, and it is difficult to choose the proper platforms for a specific usage scenario.
Therefore, it is necessary to analyze the salient characteristics of popular open-source serverless platforms and compare their performance to help developers choose suitable platforms to deploy efficient services.

\section{Platform Characteristics}
\label{sec:platform-characteristics}

Based on recent popularity, community vibrancy and feature richness, we specifically select four open-source serverless frameworks, \ie Knative, Kubeless, Nuclio and OpenFaaS, to analyze their characteristics.

\subsection{Dependency on Kubernetes}
\begin{figure}[htb]%
	\centering
	\includegraphics[width=0.45\textwidth]{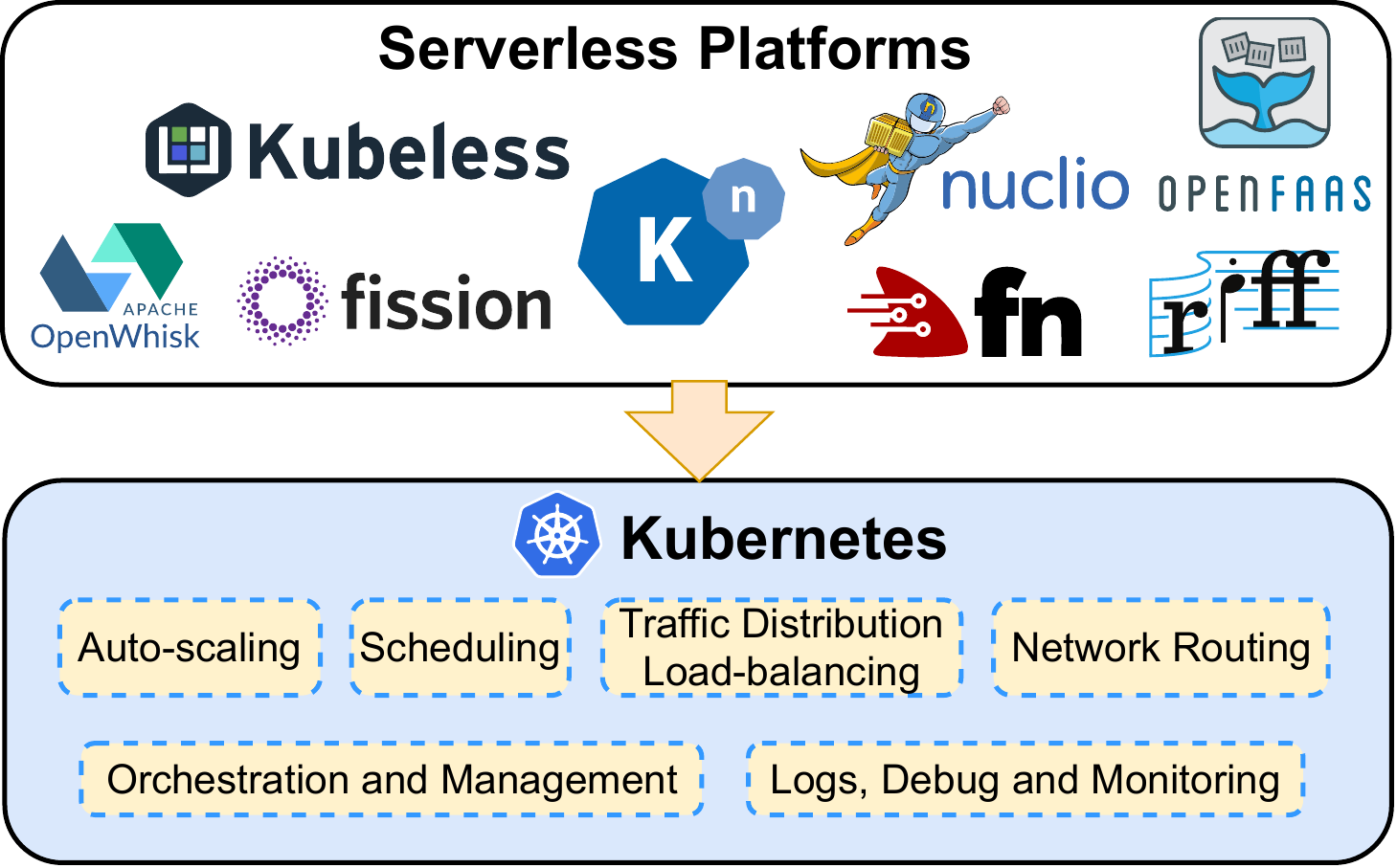}
	\caption{Serverless platforms and underlying Kubernetes services.}
	\label{fig:kubernets_svcs}
	\vspace{-4mm}
\end{figure}

Kubernetes~\cite{burns2016borg} is a portable and extensible open-source system that facilitates declarative configuration, automating deployment and management for containerized workloads. Most of open-source serverless platforms rely on Kubernetes for orchestration and management of function pods, which are the atomic deployable units in Kubernetes.
Fig.~\ref{fig:kubernets_svcs} shows the pivotal Kubernetes services that serverless platforms depend on.
These Kubernetes services are used for: (1) configuration management, (2) service discovery, (3) auto-scaling, (4) pod scheduling, (5) traffic load balancing, (6) network routing and (7) service roll-out and roll-back.

\begin{figure}[htbp]
	\vspace{-2mm}
	\centering
	\includegraphics[width=0.4\textwidth]{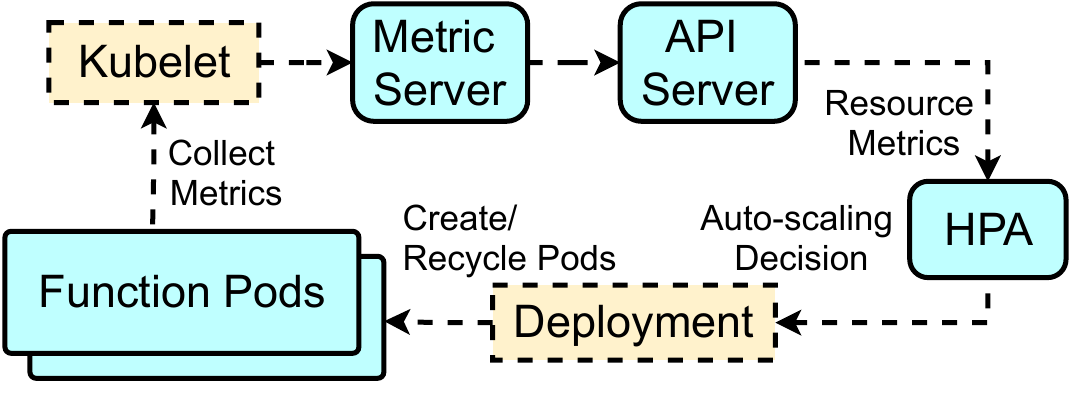}
	\caption{Horizontal pod auto-scaling framework.}
	\label{fig:hpa}
	\vspace{-4mm}
\end{figure}

Thanks to the horizontal pod auto-scaler (HPA) feature from Kubernetes, the Kubernetes-based serverless platforms support resource-based auto-scaling. The framework of HPA is shown in Fig.~\ref{fig:hpa}. The Kubelet on each node collects the resource metrics of each pod. HPA gets these metrics from the API server.
The auto-scaling threshold could be a raw value or a percentage of the pod requested amount for that resource. When the CPU or memory usage of a given function pod exceeds the threshold, HPA automatically triggers the Development controller to scale the pod number. 

\subsection{Salient Features of Serverless Platforms}

Table~\ref{table:platform-comparison} summarizes salient features of four widely-known open-source serverless platforms.

\subsubsection{Nuclio}
\begin{figure}[htb]
	\vspace{-2mm}
	\centering
	\includegraphics[width=0.45\textwidth]{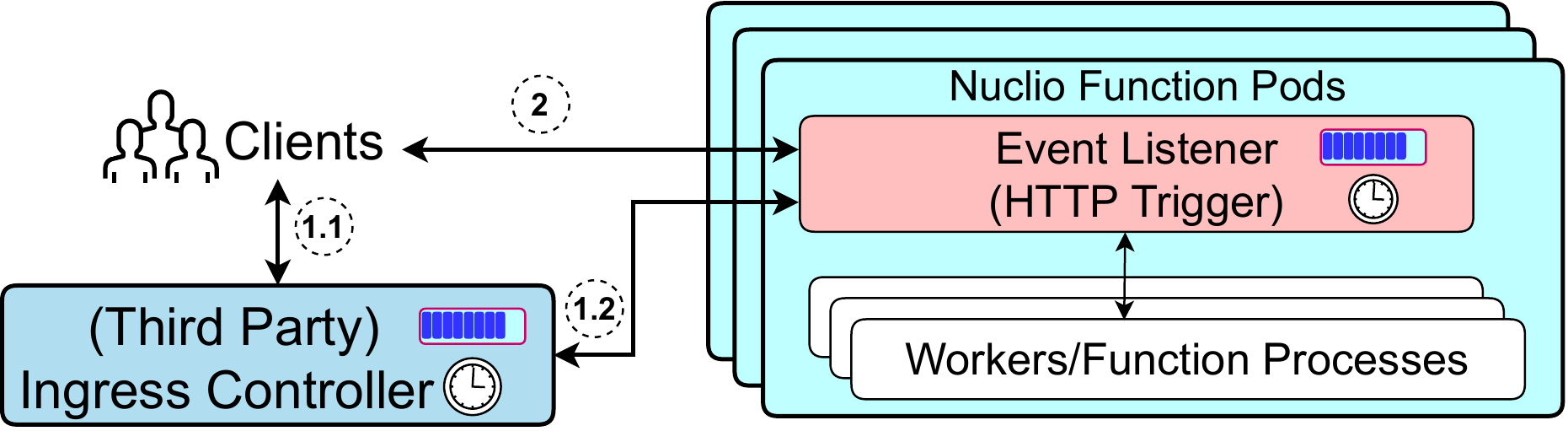}
	\caption{Nuclio framework.}
	\label{fig:nuclio_arch}
	\vspace{-4mm}
\end{figure}

The main components of Nuclio are shown in Fig.~\ref{fig:nuclio_arch}.
In each function pod, there is one event listener and multiple worker processes. The event listener receives new events and redirect them to worker processes. Multiple worker processes could work in parallel and improve the performance significantly on a multi-core worker node.
The worker process number is set to be static and specified by the configuration file.
The open-source version does not have a built-in workload-based auto-scaling feature, but the resource-based auto-scaling is supported by Kubernetes HPA.

Nuclio supports two ways to trigger functions: (1) invoking the function by name through ingress controller, which can distribute the traffic to different back-end pods according to the pre-set load balancing rule (\eg round-robin, random and least connection first) and (2) sending requests directly to function pods by NodePort, which is a unique allocated cluster-wide port for the function. In the NodePort method, incoming requests are load balanced at random by Netfilter.

\subsubsection{OpenFaaS}
\begin{figure}[htb]
	\vspace{-2mm}
	\centering
	\includegraphics[width=0.45\textwidth]{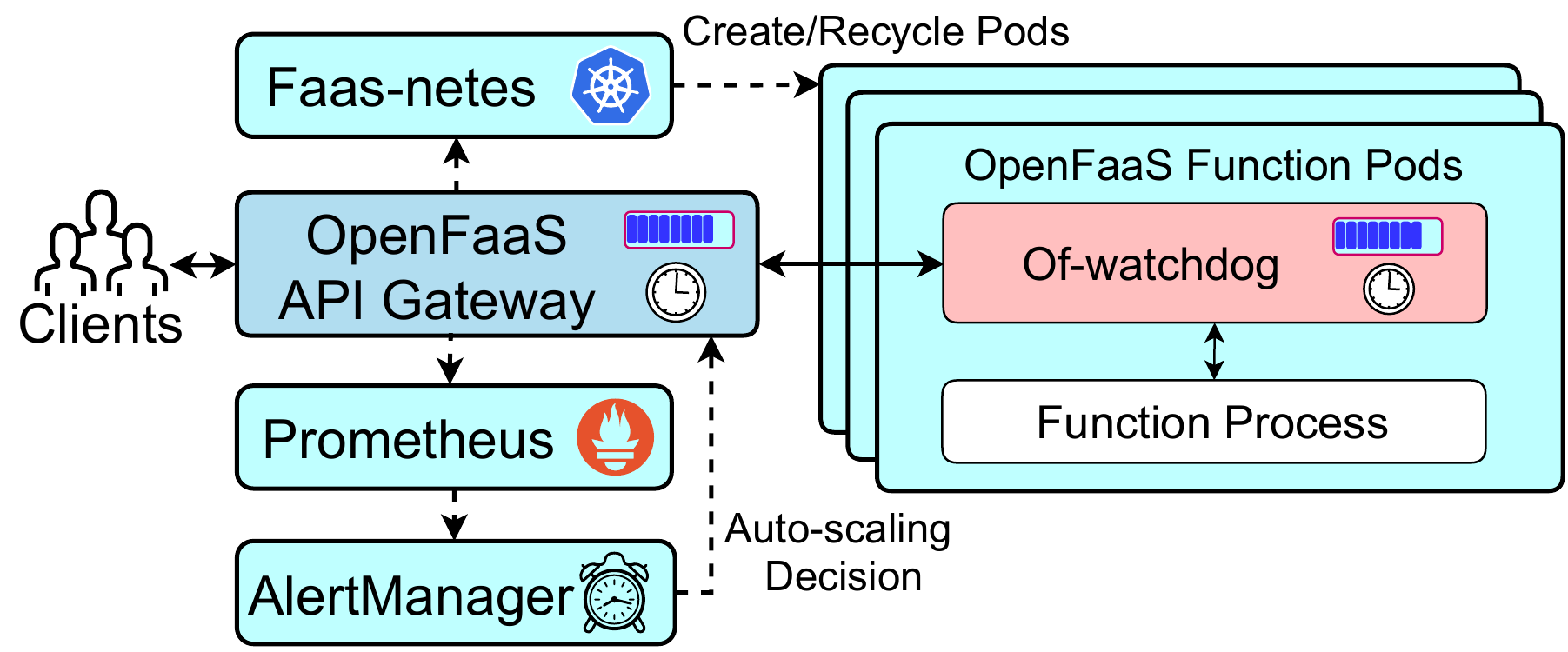}
	\caption{OpenFaaS framework.}
	\label{fig:openfaas_arch}
	\vspace{-3mm}
\end{figure}

The key components of OpenFaaS are shown in Fig.~\ref{fig:openfaas_arch}. The API gateway provides access to the functions and collects traffic metrics.
Faas-netes is the controller for managing OpenFaaS function pods. Prometheus\footnote{\url{https://prometheus.io/}} and AlertManager\footnote{\url{https://github.com/prometheus/alertmanager}} are used for auto-scaling.

Each function pod contains a single container running two type of processes, namely of-watchdog and function process. Of-watchdog is a tiny server that works as the entry-point for incoming requests and forwards them to the function process.
Of-watchdog can operate in three modes, \ie HTTP, streaming and serializing. 
In HTTP mode, the function process is forked only once at the beginning and kept warm for the entire life cycle of the function pod. 
In both the streaming and serializing mode, a new function process is forked for every request, resulting in significant cold-start latency and adverse impact on performance. 
Our evaluation results show that the throughput of the streaming or serializing mode is about $10\times$ lower than that of HTTP mode.

OpenFaaS has a built-in requests-per-second (RPS) based auto-scaling feature. Prometheus scrapes the traffic metrics from API gateway. AlertManager reads the RPS metric from Prometheus and fires an alert to the API gateway according to the auto-scaling rule defined in the configuration file. Then the API gateway handles the alert and invokes the Faas-netes to scale up or scale down function replicas.
Note that the open-source version does not support scale-to-zero feature, which is only available in the commercial version, \ie OpenFaaS Pro.

\subsubsection{Knative}

\begin{figure}[htb]
	\centering
	\includegraphics[width=0.47\textwidth]{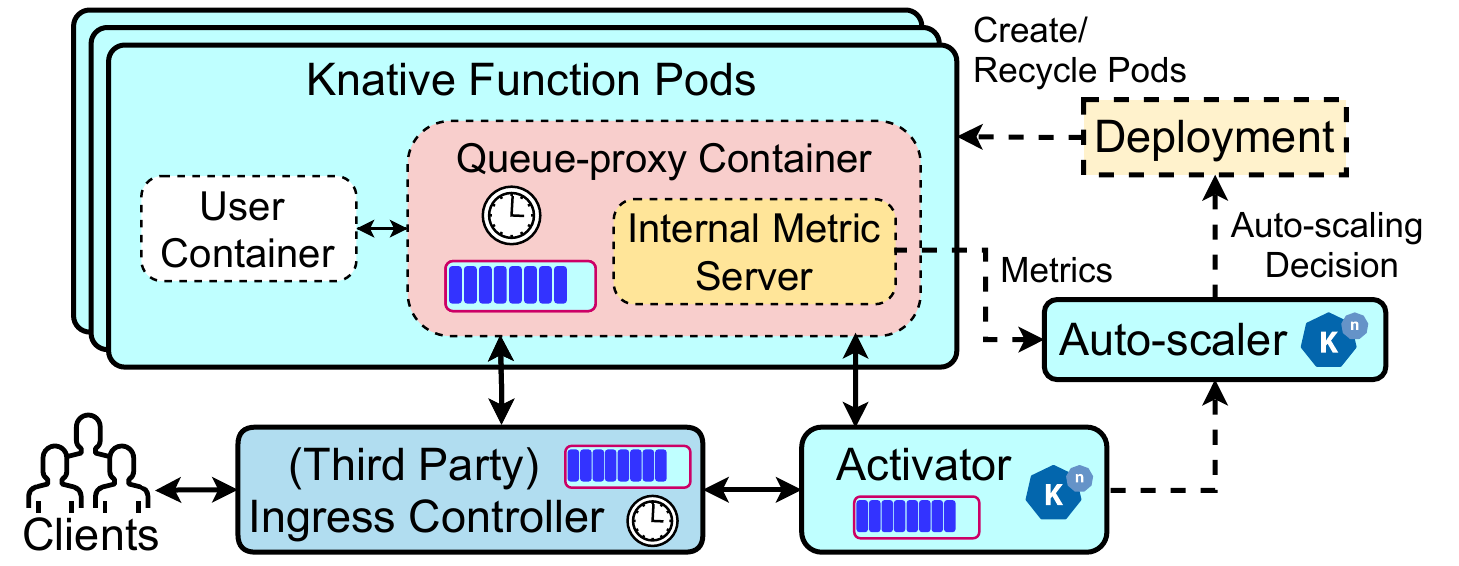}
	\caption{Knative framework.}
	\label{fig:knative-kpa}
	\vspace{-4mm}
\end{figure}

Fig.~\ref{fig:knative-kpa} shows the Knative framework.
Each function pod consists of two containers, namely queue-proxy container and function container. The queue-proxy is a sidecar container to queue incoming requests and forward them to the function container. The queue-proxy provides a buffer to handle traffic burst in spite of incurring queuing latency. In addition, the queue-proxy collects metrics and expose them via a simple HTTP server, \ie internal metric server. Multiple workers reside in the user container to process requests in parallel. 
The communication overhead between queue-proxy container and function container is higher than the process model of Nuclio and OpenFaaS, and thus results in lower performance.

The Knative built-in auto-scaling, \ie Knative pod autoscaler (KPA), supports both RPS mode and concurrency mode. The auto-scaler scrapes metrics from function pods and computes the replica number based on the auto-scaling algorithm. The deployment controller gets the auto-scaling decision and adjusts the pod number.
Knative supports scale-to-zero functionality which recycles all pods of inactive functions.
When a new request arrives for an idle function, the ingress controller redirects the request to the activator to buffer it. Then the activator triggers the autoscaler which could scale up the idle function from zero. Once the function is running again, the activator sends the buffered request to the pod. Although scale-to-zero reduces resource usage, it leads to extra cold start latency.

\subsubsection{Kubeless}
\begin{figure}[htb]
	\vspace{-2mm}
	\centering
	\includegraphics[width=0.45\textwidth]{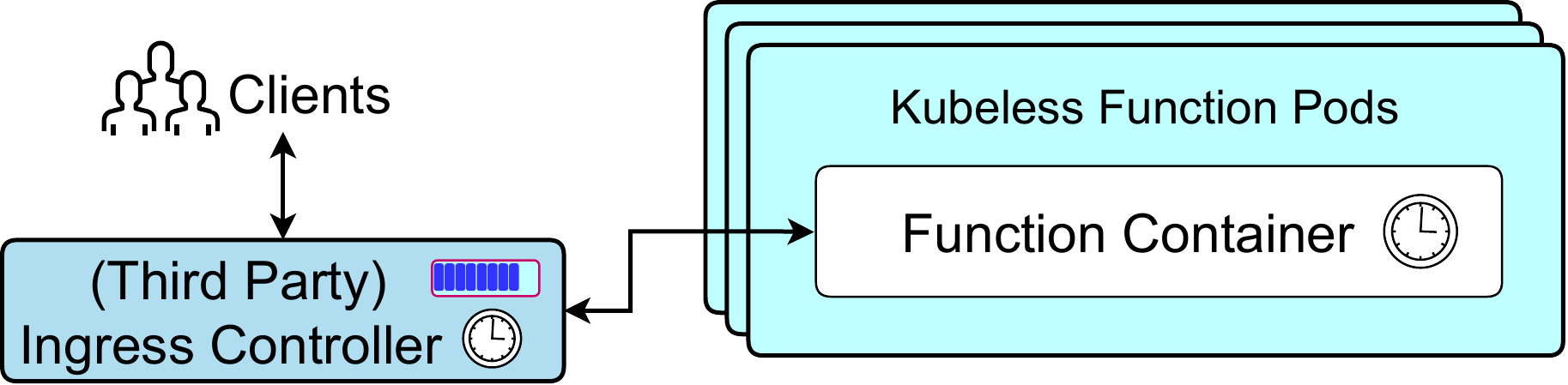}
	\caption{Kubeless framework.}
	\label{fig:kubeless_arch}
	\vspace{-4mm}
\end{figure}

Kubeless is another open-source platform built on top of Kubernetes.
Fig.~\ref{fig:kubeless_arch} describes the key components and the working model of Kubeless. 
There are several options for ingress controllers. 
We experiment with Nginx ingress controller\footnote{\url{https://kubernetes.github.io/ingress-nginx}} and Traefik ingress controller\footnote{\url{https://traefik.io}}, and opt for Traefik due to better performance. Kubeless leverages Kubernetes HPA for auto-scaling and does not support scale-to-zero.

\subsection{Service Exporting and Network Routing}

\subsubsection{Service Exporting}
The function pods are dynamic entities that can be created and destroyed at any time due to auto-scaling, failures, \etc Hence, Kubernetes provides service
as an abstraction to access the pods of the same function.
There are several ways to export services:
(1) the service could be assigned a NodePort, which is used to route the incoming traffic to the entry node and let kernel stack control load-balancing of the traffic across active pods;
(2) the API gateway/ingress controller works as the entry point and the services are exported with specific URLs.
The API gateway/ingress controller component of the serverless platform can be accessed from outside the cluster by a external public IP address.
Once the API gateway/ingress controller receives an incoming request, it determines the service for the request according to the URL, and then load-balances and routes the packet to a back-end active pod instance.

\subsubsection{Network Routing}

\begin{figure}[htb]%
	\centering
	\vspace{-2mm}
	\includegraphics[width=0.495\textwidth]{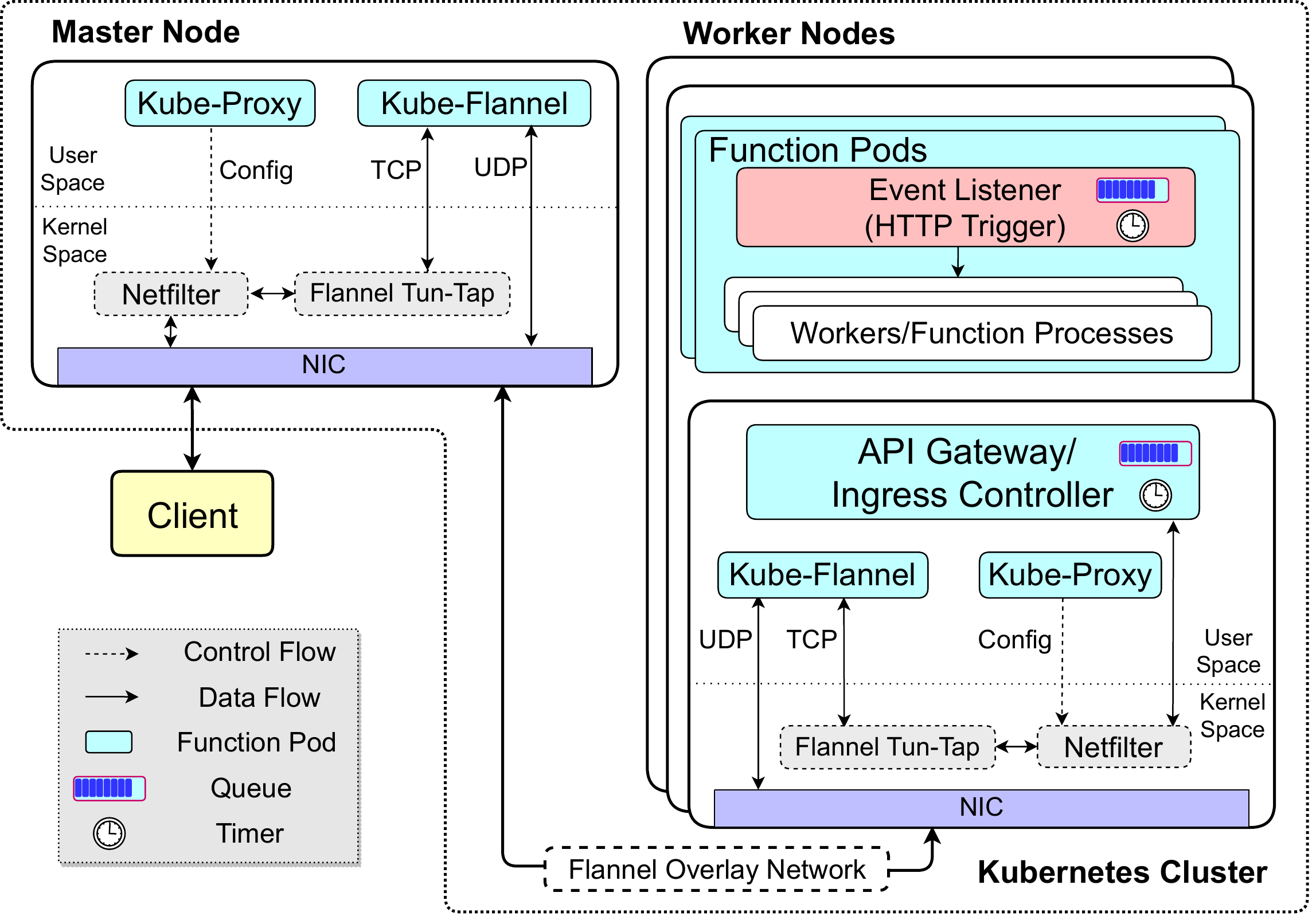}
	\caption{Network routing (Flannel mode) to export the services.}
	\label{fig:svc_export_flannel_arch}
	\vspace{-2mm}
\end{figure}

Fig.~\ref{fig:svc_export_flannel_arch} describes Flannel -- a simple Kubernetes overlay networking framework to route the traffic to function pods. 
The Kube-Proxy pod is responsible for setting up the routing and load-balancing rules, \ie the netfilter destination network address translation (DNAT) rules to change the destination IP of incoming request packets~\cite{li2017quick}.
The Kube-Flannel pod is responsible for intercepting the packets 
and performing UDP encapsulation/decapsulation for the traffic exiting/entering the physical network interface.  %

\section{Performance Evaluation}
\label{sec:evaluation}
We first compare the overall performance of different open-source serverless platforms. Based on the performance results of multiple service exporting modes and auto-scaling modes, we analyze the root cause of performance gap among different serverless platforms.

\subsection{Experimental Setup and Workload Description}
We evaluate the serverless platforms on the CloudLab testbed~\cite{cloudlab-new} consisting of one master and two worker nodes, each of them equipped with Intel CPU E5-2640v4@2.4GHz (10 physical cores), running Ubuntu 16.04.1 LTS (kernel 4.4.0-154-generic).
We build all four serverless platforms on Kubernetes (v1.20.0), using the latest version available at the time of writing\footnote{Nuclio (v1.6.1); 
OpenFaaS (v0.20.11) with HTTP mode of-watchdog;
Knative (v0.21) with Istio ingress controller (v1.8.4); 
Kubeless (v1.0.8) with Traefik ingress controller (v2.4).}.
Several serverless functions of Python 3.6 runtime are implemented.
We use %
\texttt{wrk}\footnote{\url{https://github.com/wg/wrk}}
to generate HTTP workloads for invoking serverless functions. 

\subsection{Performance}

\subsubsection{HTTP Workload}
\label{sec:httpworkload}

\begin{figure}[htpb!]
\vspace{-2mm}
\begin{subfigure}{0.5\columnwidth}
    \centering
    \includegraphics[width=\linewidth, trim=0.01cm 0.01cm 0.01cm 0.01cm, clip=true]{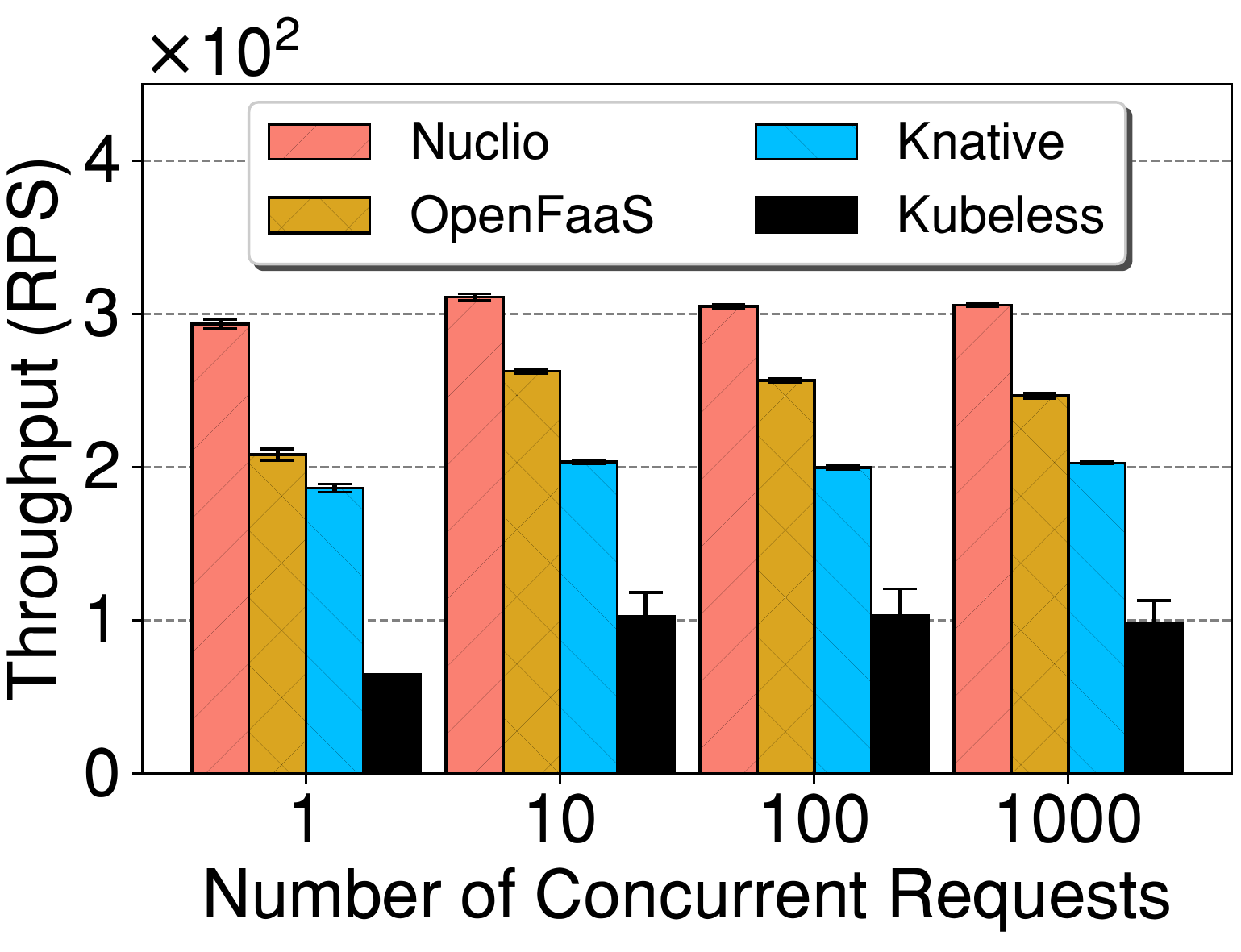}
    \caption{Throughput.}
    \label{fig:HTTPRPS}
\end{subfigure}%
\begin{subfigure}{0.5\columnwidth}\vspace{-0mm}
	\centering %
	\includegraphics[width=\linewidth, trim=0.01cm 0.01cm 0.01cm 0.01cm, clip=true]{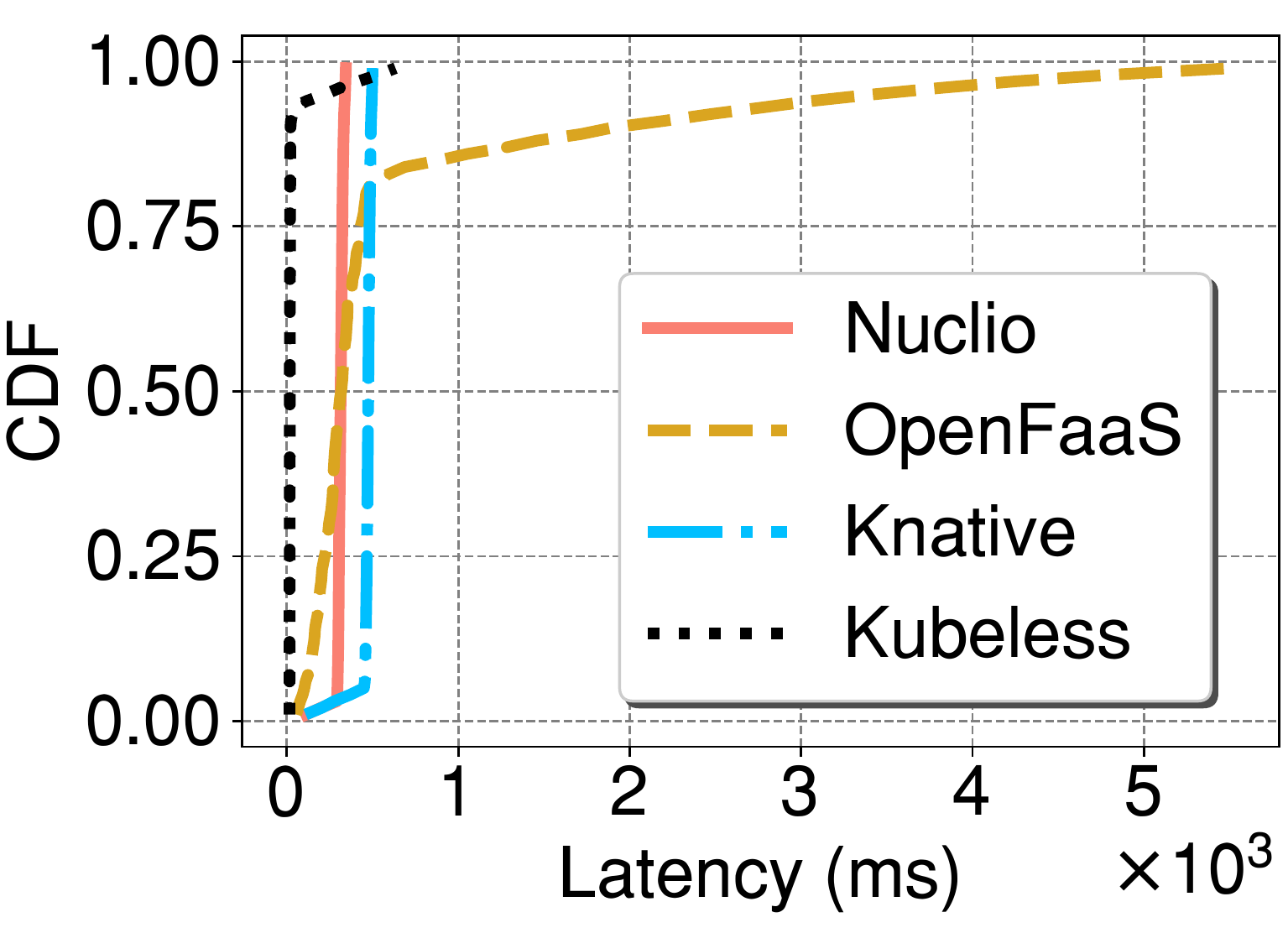}
    \caption{Latency (concurrency = 100).}
    \label{fig:HTTPCDFLatency}
\end{subfigure}%
\caption{Performance of HTTP workload function. Error bars indicate standard deviation over 20 runs.}
\label{fig:HTTPWorkload}
\vspace{-2mm}
\end{figure}

To evaluate the baseline performance of different serverless platforms, we implement a HTTP workload function that could fetch a four-byte static webpage from a local HTTP server on the master node.
For a fair comparison, we limit to %
a single instance of the function pod, disable auto-scaling and configure the same queue size and timeout parameters (50K requests, and 10s timeout) at the ingress gateway and function pod components across all the platforms. For Nuclio and Knative, we further restrict it to a single worker in one pod.
Every experiment lasts for two minutes and we measure for one minute after one-minute warm-up. The experiment is repeated for 20 times.
Fig.~\ref{fig:HTTPWorkload} shows the throughput for varying number of concurrent connections and the latency profile for concurrency level of 100. %
Nuclio has the least 99\%ile latency within 500ms,
as it allows queuing only within the function pod, while OpenFaaS and Knative can queue requests at ingress/gateway components. OpenFaaS shows heavy tail due to queuing at both the gateway and of-watchdog components.
Kubeless drops the connections at the ingress, resulting in additional retries from the client and hence lower throughput. The latency with Kubeless is lower because there is no queue inside the Kubeless function pod.

\subsubsection{Latency Breakdown of Single Request}
\begin{figure}[!htb]
	\vspace{-2mm}
    \centering
    \begin{minipage}{0.35\columnwidth}
        \centering
        \includegraphics[width=0.7\linewidth, trim=0.01cm 0.01cm 0.01cm 0.01cm, clip=true]{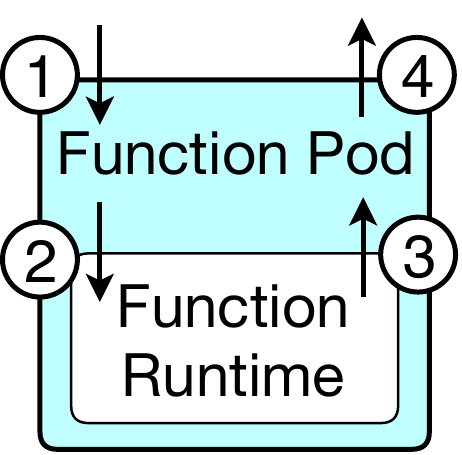}
        \label{fig:LatencyBD-1}
    \end{minipage}
    \begin{minipage}{.6\columnwidth}
        \centering
        \captionsetup{type=table} %
        \begin{tabular}{c|ccc}
        \hline
        Process & 1$\rightarrow$2 & 2$\rightarrow$3 & 3$\rightarrow$4 \\
        \hline
        Nuclio & 0.63 & 0.001 & 0.54 \\
        OpenFaaS & 1.32 & 0.001 & 0.93 \\
        Knative & 1.30 & 0.001 & 0.62 \\
        Kubeless & 4.96 & 0.001 & 2.63 \\
        \hline
        \end{tabular}
        \label{table:LatencyBD-2}
    \end{minipage}
    \caption{Latency breakdown of function execution (ms).}
    \label{fig:LatencyBD}
    \vspace{-4mm}
\end{figure}

We analyze the delay overheads incurred in processing serverless functions for different platforms. We breakdown the processing delays within the function pod. For this experiment, we use \texttt{curl} to send one request for hello-world function\footnote{It is a no operation function that returns four bytes of static text in the response.} and use \texttt{tcpdump} to capture the packets on the worker node of the function pod. We record four timestamps,
\ie (1) when the request reaches the function pod; (2) start of the function runtime; (3) end of the function runtime; (4) when the response is sent out of function pod.  
The experiment is repeated for 20 times and the average time intervals between these timestamps are shown in Fig.~\ref{fig:LatencyBD}.
In all frameworks, the actual run-time of the function (0.001ms) is the same. However, the function initiation time (time taken for request to be forwarded to the function instance) and function response delay (time taken for the response of the function to be sent out of the pod) vary. This depends on how the data is packaged and shared with the function instance. Due to forking-per-request, Kubeless incurs very high delay in forwarding the packet to the function instance. 

\subsection{Auto-scaling}
\label{eval:autoscaling}
To study the auto-scaling capabilities provided by different serverless platforms, we compare the features of both the workload-based and resource-based auto-scaling under different workload characteristics. 
We use the same HTTP workload function as in \S\ref{sec:httpworkload}.

\subsubsection{Workload-based Auto-scaling}
\label{eval:workload-based-auto-scaling}

\begin{figure}[htb!]
\vspace{-2mm}
\begin{subfigure}{0.5\columnwidth}
    \centering
    \includegraphics[width=\linewidth, trim=0.01cm 0.01cm 0.01cm 0.01cm, clip=true]{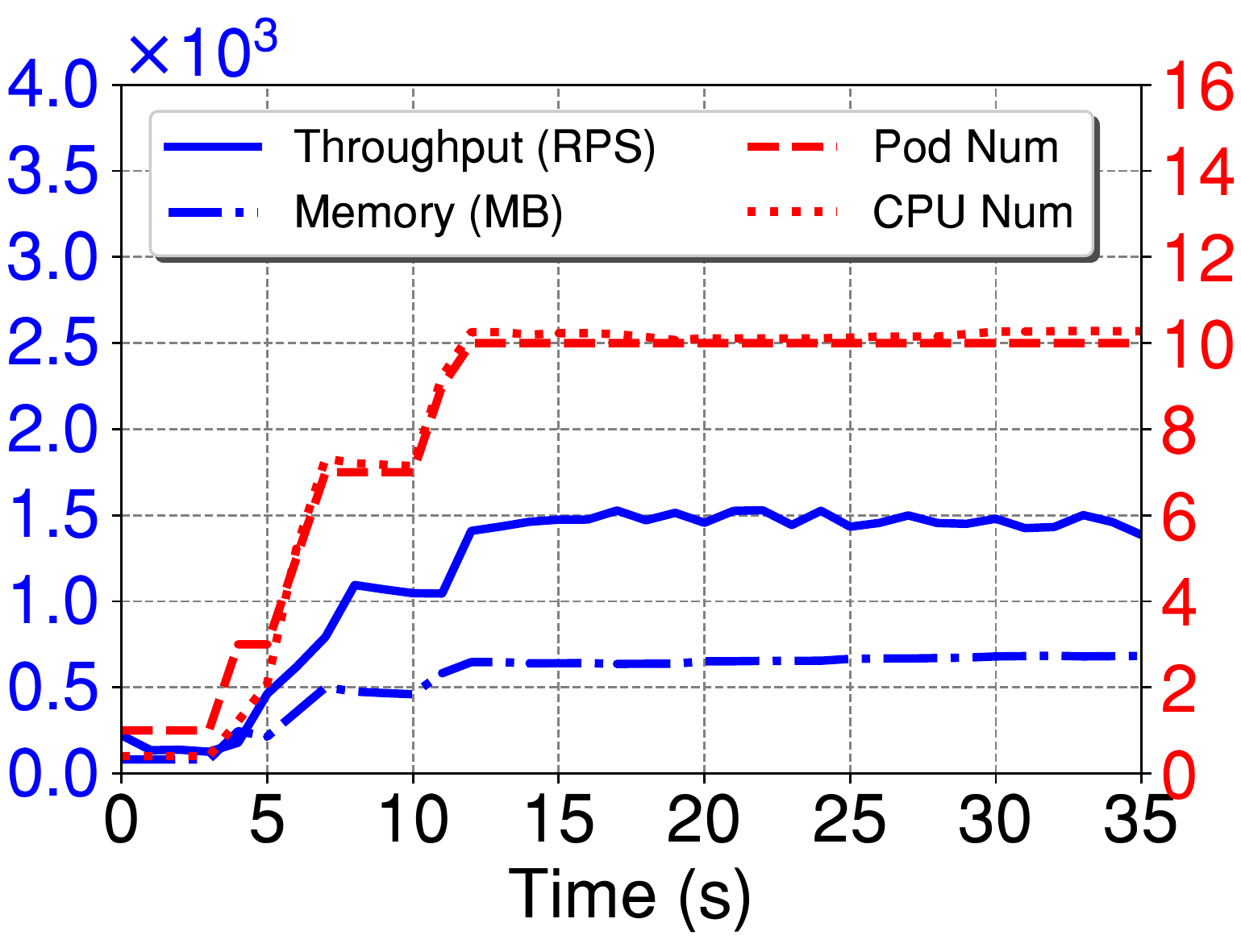}
    \caption{Knative.}
    \label{fig:Autoscaling-qps-steady-knative}
\end{subfigure}%
\begin{subfigure}{0.5\columnwidth}
	\centering
	\includegraphics[width=\linewidth, trim=0.01cm 0.01cm 0.01cm 0.01cm, clip=true]{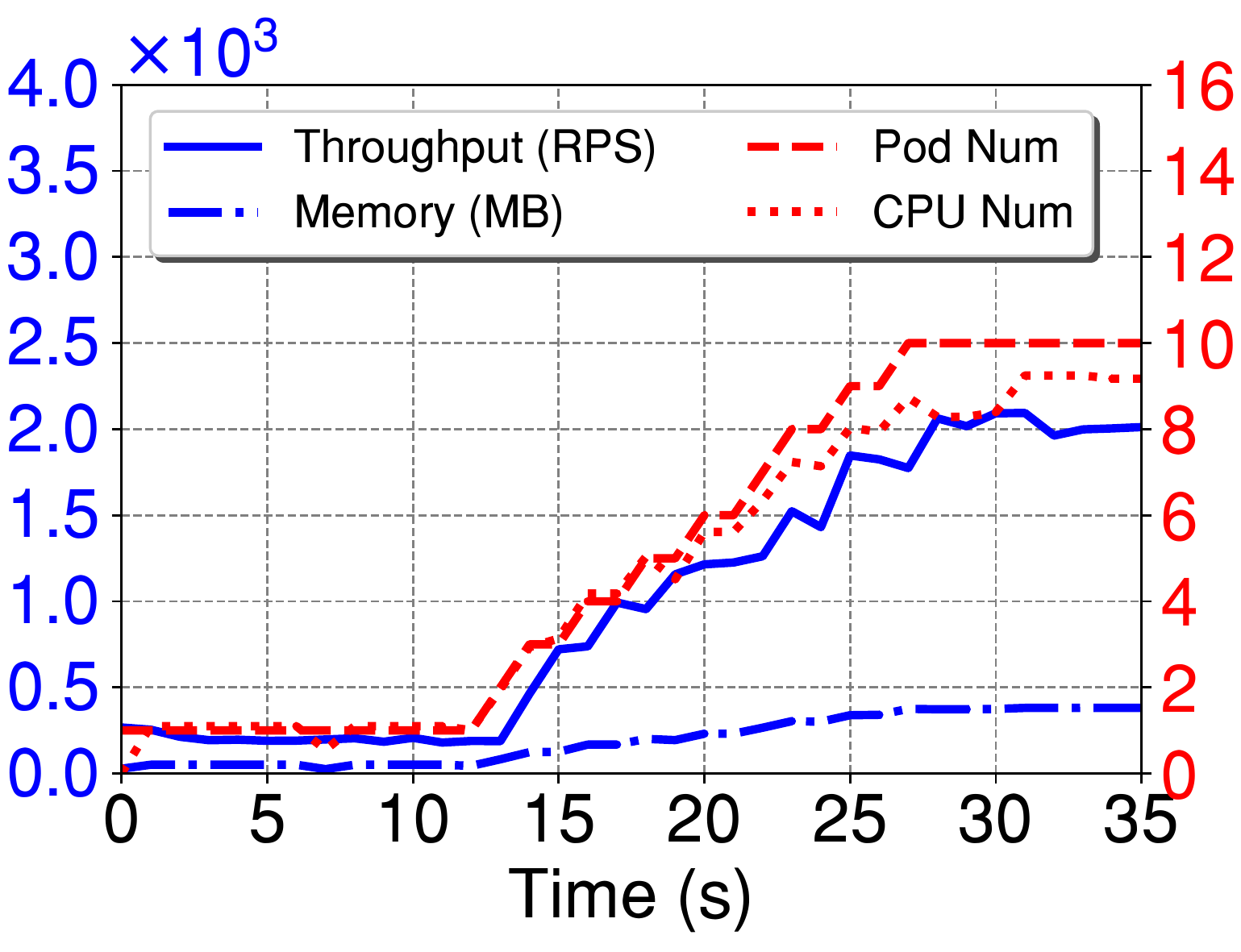}
    \caption{OpenFaaS.}
    \label{fig:Autoscaling-qps-steady-openfaas}
\end{subfigure}%
\caption{Workload-based auto-scaling.}
\label{fig:Autoscaling-qps-steady}
\vspace{-3mm}
\end{figure}

Both Knative and OpenFaaS support workload-based auto-scaling. While the workload-based auto-scaling metric of OpenFaaS is RPS, the metric in Knative is concurrency, \ie the concurrent request number.
For a fair comparison, we set equivalent auto-scaling configuration parameters for these platforms\footnote{In Knative, we set the minScale and maxScale instances as 1 and 10, target to 10, max-scale-up-rate to 100, tick interval to 2s, and stable window to 10s.
Likewise, for OpenFaaS, we set scale-factor to 10 and configure the alert-notification window to 2s, and RPS threshold to 10.}.
We use \texttt{wrk} to send a steady flow of requests (with concurrency of 100 and RPS of 100) and run the experiment for 60s.
Periodically every 2s, we monitor the number of pod instances, CPU and memory usage, and throughput.
From Fig.~\ref{fig:Autoscaling-qps-steady}, we observe that Knative scales multiple instances at a time to reach 10 instances quickly (in 12s), %
while OpenFaaS just scales up one instance at a time, taking 26s to scale up to 10 instances. 
Due to the longer process chain of auto-scaling in OpenFaaS (\ie API gateway $\rightarrow$ Prometheus $\rightarrow$ AlertManager $\rightarrow$ API gateway $\rightarrow$ Faas-netes), the scaling latency of OpenFaaS is higher than that of Knative.
Although the CPU usage for the scaled instances looks identical, the memory pressure of Knative is higher. This stems from the differences in python runtimes and proxies (\ie the queue-proxy in Knative and of-watchdog in OpenFaaS).
  
\subsubsection{Resource-based Auto-scaling}
\begin{figure}[htb!]
\vspace{-2mm}
\begin{subfigure}{0.5\columnwidth}
    \centering
    \includegraphics[width=\linewidth, trim=0.01cm 0.01cm 0.01cm 0.01cm, clip=true]{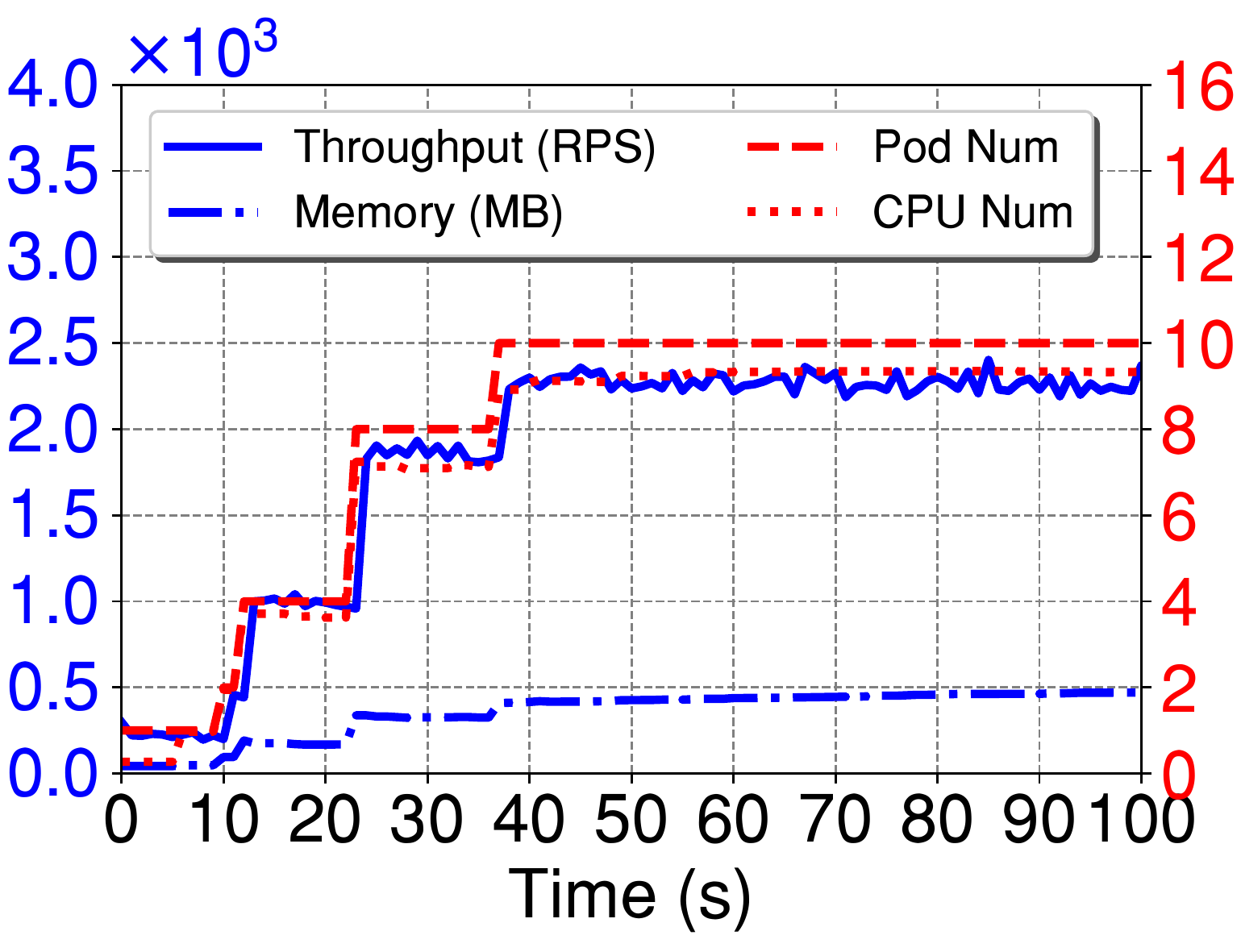}   
    \caption{Nuclio.}
    \label{fig:Autoscaling-hpa-steady-nuclio}
\end{subfigure}%
\begin{subfigure}{0.5\columnwidth}
	\centering
	\includegraphics[width=\linewidth, trim=0.01cm 0.01cm 0.01cm 0.01cm, clip=true]{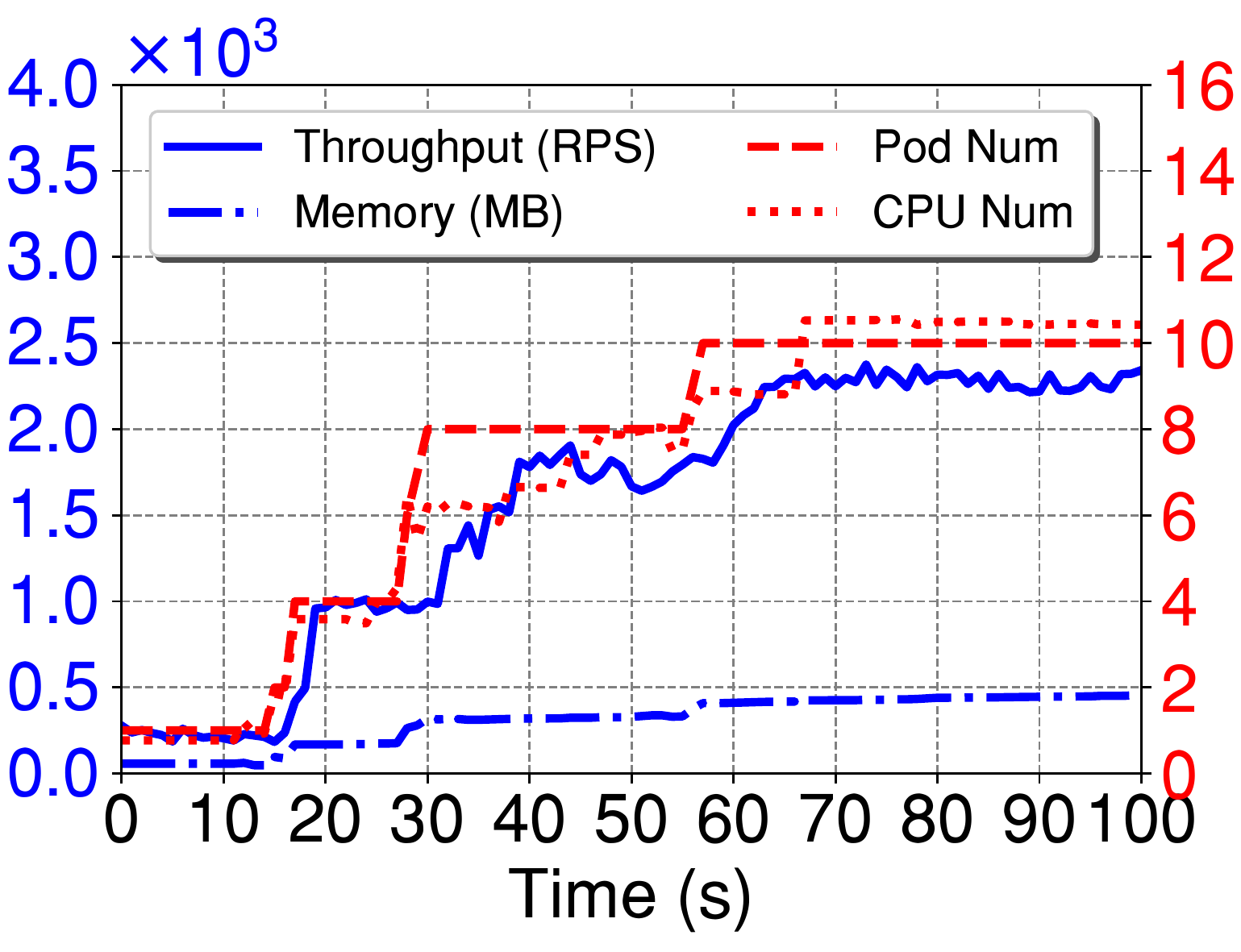}
    \caption{OpenFaaS.}
    \label{fig:Autoscaling-hpa-steady-openfaas}
\end{subfigure}
\begin{subfigure}{0.5\columnwidth}
    \centering
    \includegraphics[width=\linewidth, trim=0.01cm 0.01cm 0.01cm 0.01cm, clip=true]{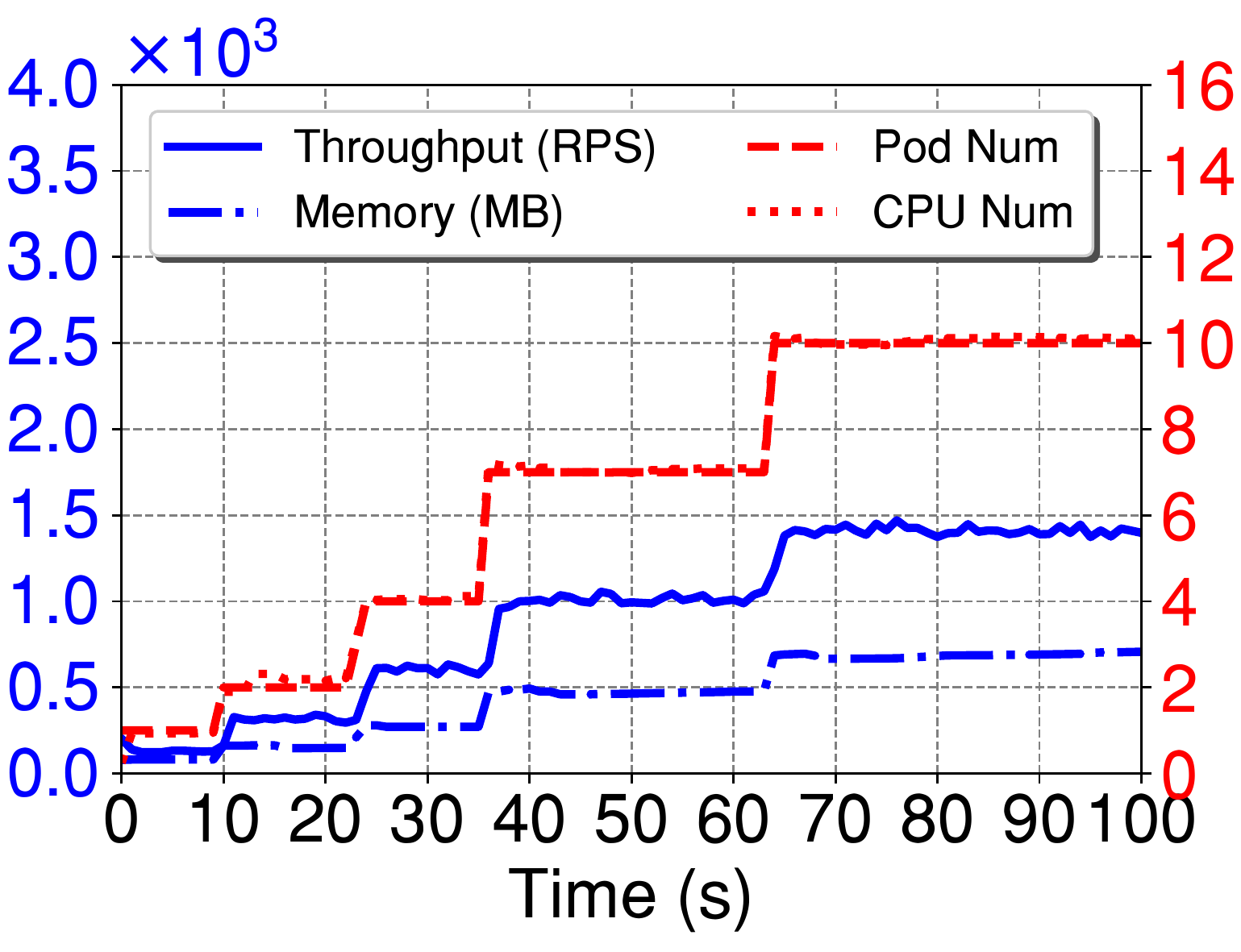}
    \caption{Knative.}
    \label{fig:Autoscaling-hpa-steady-knative}
\end{subfigure}%
\begin{subfigure}{0.5\columnwidth}
    \centering
    \includegraphics[width=\linewidth, trim=0.01cm 0.01cm 0.01cm 0.01cm, clip=true]{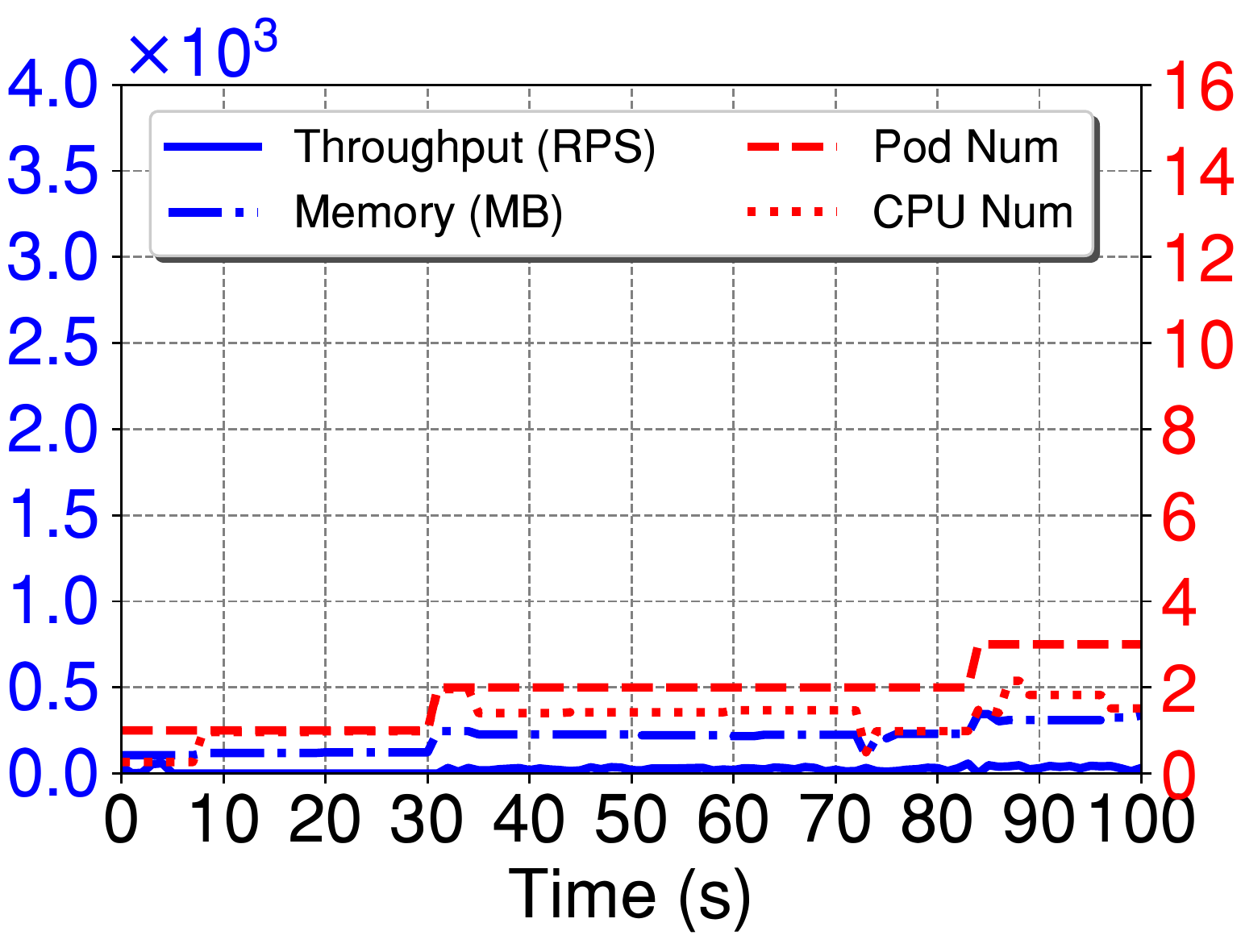}
    \caption{Kubeless.}
    \label{fig:Autoscaling-hpa-steady-kubeless}
\end{subfigure}%
\vspace{-1mm}
\caption{Resource-based auto-scaling.}
\label{fig:Autoscaling-hpa-steady}
\vspace{-6mm}
\end{figure}

All these Kubernetes-based platforms support resource-based auto-scaling. In this experiment, we use CPU usage as the metric of HPA and the CPU threshold is set to be 50\%. The other experiment configurations are the same as those in \S\ref{eval:workload-based-auto-scaling}.
As Fig.~\ref{fig:Autoscaling-hpa-steady} shows, except for Kubeless, the auto-scaling behavior is similar across all the platforms \ie auto-scaling tries to double the instances at each step until it reaches the maximum. However, the duration of each step depends on the CPU utilization factor, which in turn depends on the serverless platform specific components, such as event-listener, of-watchdog and queue-proxy. Nuclio, being relatively more CPU hungry, is able to scale more rapidly (in 40s) than Knative and OpenFaaS. %
For Kubeless, the fork-per-request and no queuing of function pods result in high latency and packet loss, which in turn contributes to lower throughput and lower CPU utilization~\cite{li2019sphinx}. Thus it leads to poor auto-scaling performance. 

\section{Insights and Future Work}
\label{sec:insights}
\subsubsection{Promising Platform}
In spite of the moderate performance compared with other open-source platforms, Knative has many useful features, such as scale-to-zero and multiple auto-scaling modes, and active community that can provide lots of help for users and developers. Thus Knative is a suitable platform for further development and innovation in serverless computing.

\subsubsection{Auto-scaling}
For current auto-scaling mechanisms, such as workload/resource-based auto-scaling, the auto-scaling metric and threshold are set by tenants. Because the tenants may not really know the runtime characteristics of their functions (\ie resource usage and execution time), they easily mis-configure the auto-scaling settings.
In addition, it is not always easy to predict the correct indicators that could show whether the current function pods are under-resourced or under-utilized.
Thus a more smart auto-scaling algorithm is needed to be designed to both properly meet the workload demand and save the resources in different scenarios.

\subsubsection{Function Startup Policy}
There are two options for function startup policy, \ie cold start and warm start. 
For the functions with low invocation rates, cold start policy could help reduce resource usage in the case of no incoming requests, but it leads to extra cold start latency. Therefore, the cold start is not appropriate for time-sensitive functions~\cite{los}.
We should carefully choose the function startup policy in accordance with the scenarios and user demands.

\subsubsection{Metric Collection}
The on-demand provisioning feature of serverless computing depends on several mechanisms, such as auto-scaling, scheduling and load balancing. All these mechanisms leverage metrics to make the decision. Many platforms, such as Knative and OpenFaaS, use scraping method to fetch metrics from pods. In our experiments, we find that the scraping method may leads to large traffic overhead when there are a large number of pods. Using sampling to only scrape a section of pods can partly decrease the overhead. However, sampling may miss out the abnormal pods and reduce the accuracy of metrics. Hence, a more efficient metric collection mechanism is worth studying further.

\subsubsection{Service Export and Network Routing}
There are many ways to export services and route incoming requests to back-end function pods, such as cluster IP, NodePort, function name/URL. All of them have both strengths and weaknesses, and should be chosen with caution.

\subsubsection{Function Chain}
It is useful to chain multiple functions for stateful workflows and complex services. How to make function chain more efficient and powerful needs to be explored in future work.

\section{Related Work}
In work~\cite{mcgrath2017serverless, wang2018peeking, understand_commodity}, the authors conducted several measurements on different cloud serverless platforms (AWS Lambda, Microsoft Azure, Google Cloud), and found the AWS to be better in terms of throughput, scalability, cold-start latency.  The work~\cite{lloyd2018serverless,cui2} investigates the different factors that influence the performance of AWS Lambda, namely the impact of the choice of language of the function, memory footprint of the function, \etc Work~\cite{mohanty2018evaluation} evaluates the performance of Fission, Kubeless and OpenFaaS serverless frameworks and characterizes the response time and the ratio of successfully completed requests for different loads. However the work fails to characterize the throughput of these platforms and accounts for the mean latency (response time) and successful responses at different load characteristics, which is debatable, without the proper consideration and configuration of the serverless platform specific configuration parameters, resulting in inaccurate results. %
Work~\cite{palade} quantitatively evaluates Apache OpenWhisk, OpenFaaS, Kubeless, and Knative platforms. The results for Kubeless are similar, but for the other platforms, we feel the presented results are inaccurate. This could be due to the usage of Kubernetes. %
In contrast, our work focuses on discerning the architectural blocks that impact the performance of Kubernetes-based open-source serverless platforms.

\section{Conclusion}
We elaborate the working models of different popular open-source serverless platforms and identify their key characteristics. 
In addition, we analyze the root causes of performance gap of different service exporting and auto-scaling modes on those platforms.
Further, several insights are proposed for future work, such as auto-scaling, service export and metric collection.

\textbf{Acknowledgment:} This work was supported by National Key Research and Development Program of China under Grant 2018YFB1800500, and US NSF grants CRI-1823270, CNS-1763929, and grants from Hewlett Packard Enterprise Co., Futurewei Technologies Inc., and China Scholarship Council. Dan Li is the corresponding author.

\bibliographystyle{IEEEtran}
\bibliography{sigproc_old}

\end{document}